\documentclass[twocolumn,showpacs,preprintnumbers,amsmath,amssymb,floatfix]{revtex4}

\usepackage{graphicx}
\usepackage{bm}

\def\mib#1{\hbox{\boldmath $#1$}}
\def\mbf#1{\hbox{\boldmath $#1$}}
\def\eq#1{Eq.\ (\ref{#1})}

\def\bq{{\mbf q}}

\def\bfnabla{{\mbf \nabla}}

\def\SM3{\Sigma N (3/2)}
\def\SN1{\Sigma N (1/2)}

\def\TS1{\hbox{}^3S_1}
\def\TD1{\hbox{}^3D_1}

\def\ah{{1 \over 2}}
\def\th{{3 \over 2}}

\begin{document}

\preprint{APS/123-QED}

\title{Faddeev Calculation of the Hypertriton using the \mib{SU_6}
Quark-Model Nucleon-Nucleon and Hyperon-Nucleon Interactions}

\author{Y. Fujiwara}
\affiliation{Department of Physics, Kyoto University, 
Kyoto 606-8502, Japan}%
 \email{fujiwara@ruby.scphys.kyoto-u.ac.jp}

\author{K. Miyagawa}%
\affiliation{Department of Applied Physics,
Okayama Science University, Okayama 700-0005, Japan}

\author{M. Kohno}%
\affiliation{Physics Division, Kyushu Dental College,
Kitakyushu 803-8580, Japan}

\author{Y. Suzuki}%
\affiliation{Department of Physics, Niigata University,
Niigata 950-2181, Japan}

\date{\today}

\begin{abstract}
Quark-model nucleon-nucleon and
hyperon-nucleon interactions by the Kyoto-Niigata group
are applied to the hypertriton calculation
in a new three-cluster Faddeev formalism using the two-cluster
resonating-group method kernels.
The most recent model, fss2, gives a reasonable
result similar to the Nijmegen soft-core model NSC89,
except for an appreciable contributions of higher partial waves. 
\end{abstract}

\maketitle



\section{Introduction}
The QCD-inspired spin-flavor $SU_6$ quark model
for the baryon-baryon interaction, proposed by
the Kyoto-Niigata group, is a unified model for the complete
baryon octet ($B_8=N$, $\Lambda$, $\Sigma$ and $\Xi$),
which has achieved very accurate descriptions
of the nucleon-nucleon ($NN$) and
hyperon-nucleon ($YN$) interactions \cite{FSS,fss2,B8B8}.
In particular, the $NN$ interaction of the most
recent model fss2 \cite{fss2} is accurate enough to compare with
the modern realistic meson-exchange models. 
These quark-model interactions can be used
for realistic calculations of few-baryon and few-cluster systems,
once an appropriate three-body equation is formulated for the
pairwise interactions described by the resonating-group
method (RGM) kernels. 
The desired three-cluster equation should be able to deal with
the non-locality and the energy-dependence intrinsically involved
in the quark-exchange RGM kernel.
Furthermore, the quark-model description of the $YN$ and 
hyperon-hyperon ($YY$) interactions in the full coupled-channel
formalism sometimes involves a Pauli forbidden state
at the quark level, which excludes the most
compact $(0s)^6$ spatial configuration,
resulting in the strongly repulsive nature
of the interactions in some particular channels.
We have recently formulated a new three-cluster equation
which uses two-cluster RGM kernels explicitly \cite{TRGM}.
This equation exactly eliminates three-cluster redundant components
by requiring the orthogonality of the total wave function
to the pairwise two-cluster Pauli-forbidden states.
The explicit energy dependence inherent
in the exchange RGM kernel is self-consistently treated.
This equation is entirely equivalent to the Faddeev equation
which uses a singularity-free $T$-matrix (which we call
the RGM $T$-matrix) generated from the two-cluster RGM kernel.

We first applied this formalism to three di-neutron
and three-alpha systems, and obtained complete equivalence between 
the Faddeev calculations and variational calculations
using the translationally invariant
harmonic-oscillator (h.o.) basis \cite{TRGM,RED}.
This formalism was also applied to the Faddeev
calculation of the three-nucleon bound state \cite{triton},
employing the off-shell $T$-matrices which are derived from the
non-local and energy-dependent RGM kernels for our
quark-model $NN$ interactions, fss2 and FSS. 
The model fss2 yields the triton binding
energy $B_t=8.519$ MeV in the 50 channel calculation,
when the $np$ interaction is employed
for all the $NN$ pairs in the isospin basis \cite{PANIC02}.
The effect of the charge dependence
of the two-body $NN$ interaction is estimated
to be $-0.19$ MeV for the triton binding energy \cite{MA89}.
This implies that our result is not overbinding in comparison with
the empirical value, ${B_t}^{\rm exp}=8.482$ MeV.
If we attribute the difference, 0.15 MeV, to the effect
of the three-nucleon force, it is by far smaller than the
generally accepted values, $0.5 \sim 1$ MeV \cite{No00},
predicted by many Faddeev calculations employing modern
realistic meson-theoretical $NN$ interactions.
The charge rms radii for $\hbox{}^3\hbox{H}$ and $\hbox{}^3
\hbox{He}$ are also correctly reproduced.
The non-local description of the short-range repulsion
in the quark model is essential to reproduce the large
binding energy and the correct size of the three-nucleon
bound state without reducing the $D$-state probability
of the deuteron.

Here we apply our quark-model $NN$ and $YN$ interactions
to the hypertriton ($\hbox{}^3_\Lambda \hbox{H}$) with
the small separation energy of the $\Lambda$-particle,
${B_\Lambda}^{\rm exp}=130 \pm 50~\hbox{keV}$.
Since the $\Lambda$-particle is far apart from
the two-nucleon subsystem, the on-shell properties
of the $\Lambda N$ and $\Sigma N$ interactions are expected to be
well reflected in this system. 
In particular, this system is very useful to determine the relative
strength of $\hbox{}^1S_0$ and $\hbox{}^3S_1$ interactions
in our framework. We will be able to finetune the quark model
interaction to the hypertriton binding energy.
This enables firmer quark model predictions
for the $\hbox{}^1S_0$ and $\hbox{}^3S_1$ phase shifts.
In fact, Ref.~\cite{MI95,MI00,No02} showed that
it is not at all trivial to bind the
hypertriton as most meson-theoretical interactions
fail to bind the hypertriton,
except for the Nijmegen soft-core potentials NSC89 \cite{NSC89},
NSC97f and NSC97e \cite{NSC97}. It is also pointed out 
in Refs.~\cite{MI95,No02} that a small admixture
of the $\Sigma NN$ components less than 1\% is
very important for this binding.

In the next section, the Faddeev equation of the hypertriton
system, using the quark-model $NN$ and $YN$ RGM kernels,
is discussed, paying a special attention
to the $\Lambda N$-$\Sigma N$ $\widetilde{T}$-matrix.
The results is given in the third section and the summary
in the last section. Appendix gives some essential points
to derive the Faddeev equation, whose solution exactly satisfies 
the orthogonality conditions.

\section{Formulation}

We start from the three-cluster equation
of the $\Lambda NN$-$\Sigma NN$ system
interacting via two-cluster RGM kernels
\begin{eqnarray}
& & P \left[ E-H_0-V^{\rm RGM}_{12}(\varepsilon_{12})
-V^{\rm RGM}_{31}(\varepsilon_{31})
-V^{\rm RGM}_{32}(\varepsilon_{32})\right]\nonumber \\
& & \qquad \times P\Psi=0\ ,
\label{form1}
\end{eqnarray}
where $E$ is the negative three-body energy measured from
the $\Lambda NN$ threshold, and the free
Hamiltonian $H_0=H^\prime+\Delta m_3$ is
composed of the kinetic-energy operators $H^\prime=h_{31}
+h_{\bar 2}$ etc. and the mass term $\Delta m_3$.
In the following, the two nucleons are numbered 1 
and 2, the $\Lambda$ or $\Sigma$ is numbered  3.  
The equation actually implies the $2 \times 2$ matrix form
and $P \Psi$ is the two-dimensional vectors composed of
the upper component with $\Lambda NN$ configuration
and the lower one $\Sigma NN$.
The mass term $\Delta m_3$ is therefore a diagonal matrix
whose matrix elements are zero for the $\Lambda NN$ channel
and $\Delta M_{\Sigma \Lambda}=M_\Sigma c^2 -M_\Lambda c^2$ for
the $\Sigma NN$ channel. Here $M_\Lambda$ and $M_\Sigma$ are
the $\Lambda$ and $\Sigma$ masses, respectively.
The RGM kernel $V^{\rm RGM}_{ij}(\varepsilon_{ij})
={V_{\rm D}}_{ij}+G_{ij}+\varepsilon_{ij} K_{ij}$ consists
of the direct potential ${V_{\rm D}}_{ij}$,
the sum of the exchange kinetic-energy and interaction
kernels, $G_{ij}=G^{\rm K}_{ij}+G^{\rm V}_{ij}$,
and the exchange normalization kernel $K_{ij}$ multiplied with
the center-of-mass energy $\varepsilon_{ij}$ of
the ${ij}$ subsystem for the relative motion.
These are also $2 \times 2$ matrices. For example,
\begin{eqnarray}
& & \varepsilon_{31}=\left(
\begin{array}{cc}
\varepsilon_{\Lambda N} & 0 \\
0 & \varepsilon_{\Sigma N} \\
\end{array}\right)
=\left(
\begin{array}{cc}
\varepsilon_{\Lambda N} & 0 \\
0 & \varepsilon_{\Lambda N-\Delta M_{\Sigma \Lambda}} \\
\end{array}\right)\ ,\nonumber \\
& & K_{31}=\left(
\begin{array}{cc}
K_{\Lambda N, \Lambda N} & K_{\Lambda N, \Sigma N} \\
K_{\Sigma N, \Lambda N}  & K_{\Sigma N, \Sigma N} \\
\end{array} \right)\ .
\label{form2}
\end{eqnarray}
The two-cluster RGM equation is expressed as
\begin{eqnarray}
\left[ \varepsilon_{31}-h_{31}-V^{\rm RGM}_{31}(\varepsilon_{31})
\right] \chi=0\ .
\label{form3}
\end{eqnarray}
The necessity of the projection operator $P$ in \eq{form1} is
related to the existence of the eigenstate
of the $\Lambda N$-$\Sigma N$ normalization kernel;
$K_{31}|u_{31}\rangle=\gamma |u_{31}\rangle$ with the
eigenvalue $\gamma=1$. This is the most compact $(0s)^6$
spin-singlet configuration with the flavor $SU_3$ quantum
number $(11)_s$; $|u_{31}\rangle
=|(0s)^6; (11)_s \hbox{}^1S_0\rangle$.
We seek for the Pauli-allowed state
of the $\Lambda NN$-$\Sigma NN$ system by diagonalization
\begin{eqnarray}
\sum_{i=1,2} |u_{3i}\rangle \langle u_{3i}|\Psi_\lambda\rangle
=\lambda |\Psi_\lambda\rangle\ ,
\label{form4}
\end{eqnarray}
in the antisymmetric model space,
$\Psi_\lambda(123)=-\Psi_\lambda(213)$,
which we denote as $\Psi_\lambda \in [11]$ in the following.
The projection operator on the Pauli-allowed space, $P$, is
defined by taking the model space spanned by the
eigenvectors with $\lambda=0$:
\begin{eqnarray}
P=\sum_{\lambda=0} |\Psi_\lambda\rangle \langle \Psi_\lambda|\ .
\label{form5}
\end{eqnarray}

The three-cluster Faddeev equation,
which is completely equivalent to \eq{form1}, is derived by
using some nice properties satisfied by $P$, which is
briefly discussed in Appendix.
The total wave function of the hypertriton system is expressed
as a superposition of two independent Faddeev
components $\psi$ and $\phi$:
$P \Psi=\psi+(1-P_{12})\phi$ with $\psi \in [11]$.
The coupled-channel Faddeev equation reads 
\begin{eqnarray}
& & \psi=G_0(E) T_{12}(E, \varepsilon_{12})(1-P_{12})\phi
\ ,\nonumber \\
& & \phi=G_0(E) \widetilde{T}_{31}(E, \varepsilon_{31})
(\psi-P_{12} \phi)\ ,
\label{form6}
\end{eqnarray}
where $P_{12}$ is the exchange operator of particle 1 and 2,
$G_0(E)=1/(E-H_0)$ is the three-body free Green function
for the negative energy $E < 0$, 
$T_{12}(E, \varepsilon_{12})$ is the $NN$ $T$-matrix
in the three-body space,
and $\widetilde{T}_{31}(E, \varepsilon_{31})$ is
the redundancy-free $\Lambda N$-$\Sigma N$ $\widetilde{T}$-matrix
in the coupled-channel formalism. These $T$-matrices are generated
from the $NN$ and $YN$ RGM kernels $V^{\rm RGM}_{NN}
(\varepsilon_{NN})$ and $V^{\rm RGM}_{YN}(\varepsilon_{YN})$,
respectively, according to the prescription
essentially given in Ref.~\cite{TRGM}.
The energy dependence involved in these kernels is treated
self-consistently by calculating the matrix elements
of the two-cluster Hamiltonian, which will be discussed
in some details.

For the $NN$ sector, $T_{12}(E, \varepsilon_{12})$ in
the $\Lambda NN$ and $\Sigma NN$ spaces are given by
the two-body $T$-matrix $t(\omega, \varepsilon)$ through
a simple replacement of the starting-energy argument:
\begin{eqnarray}
& & T_{12}(E, \varepsilon_{12})=\left(
\begin{array}{cc}
t^\Lambda_{NN} & 0 \\
0 & t^\Sigma_{NN} \\
\end{array} \right)\ ,\nonumber \\
& & t^\Lambda_{NN}=t_{NN} \left(E-{\hbar^2 \over M_N}
{\zeta_\Lambda+2 \over 4\,\zeta_\Lambda}
{\bq}^2, \varepsilon_{NN} \right)\ ,\nonumber \\
& & t^\Sigma_{NN}=t_{NN} \left(E-\Delta M_{\Sigma \Lambda}
-{\hbar^2 \over M_N}{\zeta_\Sigma+2 \over 4\,\zeta_\Sigma}
{\bq}^2, \varepsilon_{NN} \right)\ ,\nonumber \\
\label{form7}
\end{eqnarray}
where $M_N$ is the nucleon mass,
$\zeta_\beta=M_\beta/M_N$ is the mass ratio
for $\beta=\Lambda$ or $\Sigma$,
and $\bq$ is the momentum of the residual $\Lambda$ or $\Sigma$.
The $NN$ relative energy in the three-body
space, $\varepsilon_{NN}$, is determined from
\begin{eqnarray}
\varepsilon_{NN} & = & \langle P\Psi|h_{NN}
+V^{\rm RGM}_{NN}(\varepsilon_{NN})|P\Psi \rangle\ ,
\label{form8}
\end{eqnarray}
which is actually the sum of the $\Lambda NN$ and $\Sigma NN$
components. 

For the $YN$ sector, the situation is more complicated since
the $\Lambda N$-$\Sigma N$ coupling involves a complete
Pauli forbidden state $|u_{31}\rangle$ in the two-dimensional
space and the difference of the momentum dependent
starting energies in the $\Lambda N$ and $\Sigma N$ channels
is not constant because of the two types of reduced masses
between $YN$ and $N$.
In fact, the equation satisfied
by $\widetilde{T}_{31}(E, \varepsilon_{31})$ can never
be reduced to the two-cluster coupled-channel equation in the
free space. We request
\begin{eqnarray}
& & T_{31}(E, \varepsilon_{31})=V^{\rm RGM}_{31}(\varepsilon_{31})
\nonumber \\
& & \hspace{20mm} +V^{\rm RGM}_{31}(\varepsilon_{31})G_0(E)
T_{31}(E, \varepsilon_{31})\ ,\nonumber \\
& & \widetilde{T}_{31}(E, \varepsilon_{31})=
T_{31}(E, \varepsilon_{31})-(E-H_0)|u_{31}\rangle\nonumber \\
& & \times \left[E-\varepsilon_{31}-h_{\bar 2}
-\Delta m_3 \right]^{-1}\langle u_{31}|(E-H_0)\ .
\label{form9}
\end{eqnarray}
However, this equation is not actually satisfied, since
the derivation is based on the unrealistic assumptions
\begin{enumerate}
\item[1)] $\left[E-\varepsilon_{31}-h_{\bar 2}
-\Delta m_3 \right]$ is
channel independent,
\item[2)] $\left[\varepsilon_{31}-h_{31}
-V^{\rm RGM}_{31}(\varepsilon_{31})\right]|u_{31}\rangle=0$
is satisfied.
\end{enumerate}
The second condition is only approximately
satisfied since, in the strict RGM framework,
it is not possible to use the empirical internal
energies of clusters and reduced masses
for the relative kinetic-energy operators.
In fact, the correct expression for 2) is
\begin{eqnarray}
\left[\varepsilon_0-h_0
-V^{\rm RGM}_{31}(\varepsilon_0)\right]|u_{31}\rangle=0\ ,
\label{form10}
\end{eqnarray}
where $\varepsilon_0=\varepsilon_{\Lambda N}-E_{\rm int}$
with $E_{\rm int}$ being the calculated internal energies,
and $h_0$ uses the calculated reduced mass unlike $h_{31}$.
We can choose $\left(E_{\rm int}\right)_\Lambda=0$ for
the $\Lambda NN$ channel,
but $\left(E_{\rm int}\right)_\Sigma=\Delta M_{\Sigma \Lambda}$
is only approximately satisfied in our quark models, fss2 and FSS. 
This difficulty also takes place when we try to derive the basic
orthogonality condition
of the $\widetilde{T}_{31}(E, \varepsilon_{31})$
\begin{eqnarray}
\langle u_{31}| \left[\,1+G_0(\omega) \widetilde{T}_{31}
(E, \varepsilon_{31})\right]=0\ ,
\label{form11}
\end{eqnarray}
which is essential to yield the orthogonality
of the total wave function
through the second equation \eq{form6};
$\langle u_{3i}|\psi+(1-P_{12})\phi\rangle=0$ ($i=1,~2$).

Fortunately, these problems are completely solved by simply
adding a small correction term to the RGM kernel,
which is a procedure developed in Ref.~\cite{GRGM} for making
it possible to use the empirical values of the
internal energies and reduced masses in the RGM formalism.   
In Ref.~\cite{GRGM}, we have slightly modified the original
two-cluster RGM equation $\left[\varepsilon_0-h_0
-V^{\rm RGM}_{31}(\varepsilon_0)\right]\chi=0$ and
considered the following RGM equation
in the OCM (orthogonality condition model) form:
\begin{eqnarray}
& & \Lambda \left[ \varepsilon-h^{\rm exp}_0
-V_{\rm RGM}(\varepsilon_0)
\right] \Lambda \chi=0\ .
\label{form12}
\end{eqnarray}
From here on, we omit the subscript ${31}$ or $YN$ as 
much as possible, in order to simplify the notations.
For example, $V_{\rm RGM}(\varepsilon_0)
=V^{\rm RGM}_{31}(\varepsilon_0)$ 
and $\Lambda=\Lambda_{31}=1-|u\rangle \langle u|$
with $|u\rangle=|u_{31}\rangle$ in \eq{form12}.
Furthermore, $\varepsilon=\varepsilon_{\Lambda N}
-E^{\rm exp}_{\rm int}$ in \eq{form12} uses the empirical
internal energy, $E^{\rm exp}_{\rm int}=\Delta m_3$, and
the relative kinetic-energy
operator, ${h_0}^{\rm exp}=h_{31}=
-(\hbar^2/2\mu^{\rm exp})\bfnabla^2$, uses the empirical reduced
mass, $\mu^{\rm exp}=\mu^{\rm exp}_{\Lambda N}$ or
$\mu^{\rm exp}_{\Sigma N}$.
On the other hand, we need to
use $\varepsilon_0$ with the calculated internal energies
in the RGM kernel.
It is shown in Ref.~\cite{GRGM} that \eq{form12} is
converted to \eq{form3} by simply adding
$\Delta G$ to $V_{\rm RGM}(\varepsilon_0)$; i.e.,
$V^{\rm RGM}_{31}(\varepsilon_{31})=V_{\rm RGM}(\varepsilon_0)
+\Delta G$. The explicit expression of $\Delta G$ is
given in the paper (or $\Delta \varepsilon=0$ case
in \eq{form16} below).
We use the same idea to eliminate the channel dependence
of $\left[E-\varepsilon_{31}-h_{\bar 2}-\Delta m_3 \right]$ in 1).
Let us start from the off-shell extension
of the $\Lambda N$-$\Sigma N$ coupled-channel RGM equation
in the following OCM form:
\begin{eqnarray}
& & \Lambda \left[\,\omega-{h_0}^{\rm exp}
- V_{\rm RGM}(\varepsilon_0)
\,\right] \Lambda \chi+|u\rangle (\omega-\varepsilon)
\langle u|\chi\rangle=0\ .\nonumber \\
\label{form13}
\end{eqnarray}
The solution of this equation satisfies the orthogonality,
$\langle u|\chi\rangle=0$ for $\omega \neq \varepsilon$.
The various energies in \eq{form13} are usually channel
dependent; namely,
if we use the label $\beta=\Lambda$ or $\Sigma$ to specify
the $\Lambda N$ or $\Sigma N$ channel, the diagonal
matrix elements of these energies are given by
\begin{eqnarray}
\omega
& = & E-{\hbar^2 \over 2M_N}{\zeta_\beta+2 \over \zeta_\beta+1}
\bq^2-E^{\rm exp}_{\rm int}, \nonumber \\
\varepsilon & = & \varepsilon_{\Lambda N}
-{\hbar^2 \over 2M_N}\left( {\zeta_\beta+2 \over \zeta_\beta+1}
-{\zeta_\Lambda+2 \over \zeta_\Lambda+1}\right) \bq^2
-E^{\rm exp}_{\rm int}\ ,\nonumber \\
\varepsilon_0 & = & \varepsilon_{\Lambda N}
-E_{\rm int}\ ,
\label{form14}
\end{eqnarray}
where $E$ is the negative three-body energy,
$\bq$ this time is the momentum of the residual nucleon.
We can prove that \eq{form13} is equivalent to the following 
Schr{\"o}dinger-type RGM equation:
\begin{eqnarray}
\omega \chi=\left[\,{h_0}^{\rm exp}
+V^{\rm mod}_{\rm RGM}(\varepsilon)\,\right] \chi\ ,
\label{form15}
\end{eqnarray}
where a newly defined RGM
kernel $V^{\rm mod}_{\rm RGM}(\varepsilon)$ is
given by
\begin{eqnarray}
V^{\rm mod}_{\rm RGM}(\varepsilon)
& = & V_{\rm RGM}(\varepsilon_0)+\Delta G\ ,\nonumber \\
\Delta G & = & \Lambda \left(\Delta E_{\rm int}+\Delta \varepsilon
+\Delta h_0 \right) \Lambda \nonumber \\
& & -\left(\Delta E_{\rm int}+\Delta \varepsilon
+\Delta h_0 \right)\ ,\nonumber \\
\Delta E_{\rm int} & = & E^{\rm exp}_{\rm int}-E_{\rm int}
\ ,\quad \Delta h_0 = {h_0}^{\rm exp}-h_0\ ,\nonumber \\
\Delta \varepsilon & = & {\hbar^2 \over 2M_N}
\left( {1 \over \zeta_\beta+1}
-{1 \over \zeta_\Lambda+1}\right) \bq^2\ .
\label{form16}
\end{eqnarray}
The RGM $T$-matrix $\widetilde{T}_{31}(E, \varepsilon_{31})$ is
therefore formulated for this modified RGM
kernel $V^{\rm mod}_{\rm RGM}(\varepsilon)$.
By repeating the same process as to derive \eq{form9}
with respect to $V^{\rm mod}_{\rm RGM}(\varepsilon)$,
we can find that \eq{form9} (and also \eq{form11}) is just valid
if we replace (see Appendix for details)
\begin{eqnarray}
& & V^{\rm RGM}_{31}(\varepsilon_{31})
\rightarrow V^{\rm mod}_{\rm RGM}(\varepsilon)\ ,\nonumber \\
& & \left[E-\varepsilon_{31}-h_{\bar 2}-\Delta m_3 \right]^{-1}
\rightarrow
\left[E-\varepsilon_{\Lambda N}-\left(h_{\bar 2}\right)_\Lambda
\right]^{-1}\ .\nonumber \\
\label{form17}
\end{eqnarray}

In order to determine $\varepsilon_{\Lambda N}$ or $\varepsilon_0$,
we approximate $\omega$ in \eq{form15} as
\begin{eqnarray}
& & \omega
=\left( \begin{array}{cc}
E-{\hbar^2 \over 2M_N}{\zeta_\Lambda+2 \over \zeta_\Lambda+1}\bq^2
  & 0 \\
0 & E-{\hbar^2 \over 2M_N}{\zeta_\Sigma+2 \over \zeta_\Sigma+1}\bq^2
-\Delta M_{\Sigma \Lambda} \\
\end{array}
\right) \nonumber \\
& & \qquad \longrightarrow
\left( \begin{array}{cc}
\varepsilon_{\Lambda N} & 0 \\
0 & \varepsilon_{\Sigma N} \\
\end{array}
\right)\ \ ,
\label{form18}
\end{eqnarray}
with keeping the relationship $\varepsilon_{\Sigma N}
=\varepsilon_{\Lambda N}-\Delta M_{\Sigma \Lambda}$ in the free
space, and first calculate a $\Lambda N$-$\Sigma N$ averaged value
\begin{equation}
\varepsilon_{YN}=\langle P \Psi|\,h^{\rm exp}_0
+V^{\rm mod}_{\rm RGM}(\varepsilon)\,|
P \Psi \rangle\ .
\label{form19}
\end{equation}
Then, from $\varepsilon_{YN}=\varepsilon_{\Lambda N}
\,\langle \Psi|\Psi\rangle^\Lambda
+\varepsilon_{\Sigma N}\,\langle \Psi|\Psi\rangle^\Sigma$
and $\Delta M_{\Sigma \Lambda}=\varepsilon_{\Lambda N}
-\varepsilon_{\Sigma N}$, we find
\begin{eqnarray}
\varepsilon_{\Lambda N} & = & \varepsilon_{YN}
+\Delta M_{\Sigma \Lambda}\,P_\Sigma\ ,\nonumber \\
\varepsilon_{\Sigma N} & = & \varepsilon_{YN}
-\Delta M_{\Sigma \Lambda}\,(1-P_\Sigma)\ ,
\label{form20}
\end{eqnarray}
where $P_\Sigma=\langle \Psi |\Psi \rangle^\Sigma$ is
the probability of the $\Sigma NN$ component admixed
in the hypertriton wave function.

\begin{table}[htb]
\caption{
The channel quantum numbers of the hypertriton included
in 15-channel Faddeev calculation with $S$- and $D$-waves.
For the $NN$-$Y$ channels, $I$ and $j$ are coupled to the
total angular-momentum $\protect\ah^+$ and the isospin coupling
is $[(\protect\ah \protect\ah)t;t]0$.
For the $YN$-$N$ channels, these are $[I_1 j_2]\protect\ah$ and
$[(t\protect\ah)\protect\ah;\protect\ah]0$.
The isospin value $t$ specifies the
hyperon species; i.e., $Y=\Lambda$ for $t=0$ and $Y=\Sigma$
for $t=1$.}
\label{table1}
\begin{center}
\renewcommand{\arraystretch}{1.4}
\setlength{\tabcolsep}{5mm}
\begin{tabular}{cccc}
\hline
$Y$-$NN$ & $\hbox{}^{2s+1}\lambda_I$
 & $\left(\ell \protect\ah\right)j$
 & $t$ \\
\hline
1  &  $\hbox{}^3S_1$ & $(S\protect\ah)\protect\ah$  & 0 \\
2  &  $\hbox{}^3S_1$ & $(D\protect\ah)\protect\th$ & 0 \\
3  &  $\hbox{}^3D_1$ & $(S\protect\ah)\protect\ah$  & 0 \\
4  &  $\hbox{}^3D_1$ & $(D\protect\ah)\protect\th$ & 0 \\
5  &  $\hbox{}^1S_0$ & $(S\protect\ah)\protect\ah$  & 1 \\
\hline
$N$-$YN$ & $\hbox{}^{2s_1+1}{\left(l_1\right)}_{I_1}$
 & $\left(\ell_2 \ah\right)j_2$ & $t$ \\
\hline
1  &  $\hbox{}^3S_1$ & $(S\protect\ah)\protect\ah$  & 0~(1) \\
2  &  $\hbox{}^3S_1$ & $(D\protect\ah)\protect\th$ & 0~(1) \\
3  &  $\hbox{}^3D_1$ & $(S\protect\ah)\protect\ah$  & 0~(1) \\
4  &  $\hbox{}^3D_1$ & $(D\protect\ah)\protect\th$ & 0~(1) \\
5  &  $\hbox{}^1S_0$ & $(S\protect\ah)\protect\ah$  & 0~(1) \\
\hline
\end{tabular}
\end{center}
\end{table}

\section{Result}

The angular-momentum states of the $NN$-$Y$ channel is
specified by $|[(\lambda s)I,(\ell \ah)j]\ah;
[\left(\ah \ah\right)t,t]0 \rangle$, where $(\lambda s)I$ stands
for the two-nucleon state, $\hbox{}^{2s+1}\lambda_I$,
and $t$ is the isospin value $t=0$ for $\Lambda$ and $t=1$ for
$\Sigma$. Due to the antisymmetric property of the two
nucleons, $(-)^{\lambda+s+t}=-1$, we find that
the $\hbox{}^3E$ and $\hbox{}^1O$ states contribute
to the $\Lambda NN$ channel and the $\hbox{}^1E$
and $\hbox{}^3O$ states contribute to the $\Sigma NN$ channel.
For the $YN$-$N$ channel, the angular-momentum
states are specified by $|[(\ell_1 s_1)I_1,(\ell_2 \ah)j_1]\ah;
[\left(t \ah\right)\ah,\ah]0 \rangle$.
Since the isospin of the hypertriton is zero, only the
isospin $T=1/2$ sector of the $\Lambda N$-$\Sigma N$ interaction
contributes to the hypertriton calculation. All the partial
waves of the orbital angular-momentum are possible for each
of the $\Lambda NN$ and $\Sigma NN$ channels, which makes 
the number of channels for a particular partial-wave truncation
just three-times larger than in the triton Faddeev calculation.
The hyperon species of the $YNN$ channels are uniquely
specified by the isospin value $t=0$ or 1.
For the orbital part, the parity conservation
requires $(-1)^{\lambda+\ell}=(-1)^{\ell_1+\ell_2}=1$.
The channel truncation is specified by the maximum value
of the total angular momenta of the pairwise baryons,
$I$ and $I_1$, which we denote $J$.
As an example, all the channels of the standard 15-channel
calculation with $S$- and $D$-waves are listed
in Table \ref{table1}.

\begin{table}[htb]
\caption{Results of the hypertriton Faddeev
calculations by fss2 and FSS.
The momentum descritization points are $n_1$-$n_2$-$n_3$=10-10-5
with $q_1$-$q_2$-$q_3$=1-3-6 ${\rm fm}^{-1}$.
The calculated deuteron binding energy
is $\varepsilon_d=2.2247$ MeV for fss2 and 2.2561 MeV for FSS
($\varepsilon^{\rm exp}_d=2.2246$ MeV).
The heading $E$ is the hypertriton
energy measured from the $\Lambda NN$ threshold; $B_\Lambda$ is
the $\Lambda$ separation energy; $\varepsilon_{NN}$
($\varepsilon_{\Lambda N}$) is the $NN$ ($\Lambda N$) expectation
value determined self-consistently;
and $P_\Sigma$ is the $\Sigma NN$ probability in percent.
The norm of redundant components,
$N^2_{\rm red}=\langle u_{3i}|P \Psi \rangle^2$ ($i=1,~2$),
is less than $10^{-9}$.}
\label{table2}
\renewcommand{\tabcolsep}{0.3pc} 
\renewcommand{\arraystretch}{1.2} 
\begin{tabular}{@{}ccccccc}
\hline
model & No. of & $E$ & $B_\Lambda$
& $\varepsilon_{NN}$ & $\varepsilon_{\Lambda N}$
& $P_\Sigma$ \\
      & channels & (MeV) & (keV) & (MeV) & (MeV) & $(\%)$ \\
\hline
 &   6 ($S$)  & $-2.362$ & 137 & $-1.815$ & 3.548 & 0.450 \\
 &  15 ($SD$) & $-2.423$ & 198 & $-1.762$ & 5.729 & 0.652 \\
 &  30 ($J \le 1$) & $-2.403$ & 178 & $-1.786$ & 5.664 & 0.615 \\
fss2 & 54 ($J \le 2$) & $-2.498$ & 273 & $-1.673$ & 5.974 & 0.777 \\
 &  78 ($J \le 3$) & $-2.510$ & 285 & $-1.660$ & 6.014 & 0.800 \\
 & 102 ($J \le 4$) & $-2.513$ & 288 & $-1.658$ & 6.022 & 0.804 \\
 & 126 ($J \le 5$) & $-2.514$ & 289 & $-1.657$ & 6.024 & 0.805 \\
 & 150 ($J \le 6$) & $-2.514$ & 289 & $-1.657$ & 6.024 & 0.805 \\
\hline
 &   6 ($S$)  & $-2.910$ & 653 & $-1.309$ & 3.984 & 1.022 \\
 &  15 ($SD$) & $-2.967$ & 710 & $-1.433$ & 6.171 & 1.200 \\
 &  30 ($J \le 1$) & $-2.947$ & 691 & $-1.427$ & 6.143 & 1.191 \\
FSS & 54 ($J \le 2$) & $-3.121$ & 865 & $-1.323$ & 6.467 & 1.348 \\
 &  78 ($J \le 3$) & $-3.128$ & 872 & $-1.320$ & 6.480 & 1.357 \\
 & 102 ($J \le 4$) & $-3.134$ & 877 & $-1.317$ & 6.488 & 1.360 \\
 & 126 ($J \le 5$) & $-3.134$ & 878 & $-1.317$ & 6.488 & 1.361 \\
 & 150 ($J \le 6$) & $-3.134$ & 878 & $-1.317$ & 6.489 & 1.361 \\
\hline
\end{tabular}
\end{table}

For the numerical calculation, we discretize the continuous
momentum variable $p$ (or $q$) for the Jacobi coordinate vectors,
using the Gauss-Legendre $n_1$- (or $n_2$-) point
quadrature formula, for each of the three intervals
of 0 - 1 $\hbox{fm}^{-1}$,
1 - 3 $\hbox{fm}^{-1}$ and 3 - 6 $\hbox{fm}^{-1}$. 
The small contribution from the intermediate integral
over $p$ beyond $p_0=6~\hbox{fm}^{-1}$ in the two-body $T$-matrix
calculation is also taken into account by using
the Gauss-Legendre $n_3$-point quadrature formula through the
mapping $p=p_0+{\rm tan}\left\{\pi(1+x)/4\right\}$.
[These $n_3$ points for $p$ are not included
for solving the Faddeev equation \eq{form6},
since it causes a numerical inaccuracy for the interpolation.]
The momentum region $q=$ 6 $\hbox{fm}^{-1}$ - $\infty$ is
also discretized by the $n_3$-point formula just
as in the $p$ discretization case.
We take $n_1$-$n_2$-$n_3$=10-10-5 as is used for the triton
Faddeev calculation in Refs.~\cite{triton,PANIC02},
for which well converged results are obtained.
The partial-wave decomposition of
the two-cluster RGM kernel is carried out numerically using
the Gauss-Legendre 20-point quadrature formula.
The modified spline interpolation technique
developed in Ref.~\cite{GL82a} is employed for simplifying
the treatment of the rearrangement of the Jacobi momentum 
coordinates.
The Faddeev formalism with two identical particles or clusters
is discussed in Ref.\ \cite{BE9L},
together with some formulas for calculating the matrix
elements of the two-cluster Hamiltonian.
For the diagonalization of the large non-symmetric matrix
appearing in solving Faddeev equations,
the Arnoldi-Lanczos algorithm developed in the ARPACK
subroutine package \cite{AR96} is very useful.

Table \ref{table2} shows the results of the Faddeev calculations
using fss2 and our previous model FSS.
In the 15-channel calculation including the $S$ and $D$ waves
of the $NN$ and $YN$ interactions,
we have already obtained $B_\Lambda=-\varepsilon_d
-E(\hbox{}^3_\Lambda \hbox{H}) \sim 200$ keV for fss2.
The convergence with the extension to the higher partial waves
is very rapid, and the total angular-momentum
of the baryon pairs with $J \leq 4$ is
good enough for 1 keV accuracy.
As for the converged $B_\Lambda$ values
with 150-channel $\Lambda NN$ and $\Sigma NN$ configurations,
we obtain $B_\Lambda=289$ keV and
the $\Sigma NN$ component $P_\Sigma=0.80\,\%$ for the
fss2 prediction,
and $B_\Lambda=878$ keV and $P_\Sigma=1.36\,\%$ for FSS.
Table \ref{table3} shows the correlation
between the $\Lambda$ separation
energy $B_\Lambda$ and the $\hbox{}^1S_0$ and $\hbox{}^3S_1$
effective range parameters of FSS,
fss2 and NSC89 $\Lambda N$ interactions.
Although all of these $\Lambda N$ interactions reproduce
the low-energy $\Lambda N$ total cross section data
below $p_\Lambda \sim 300~\hbox{MeV}/c$ within the experimental
error bars, our quark-model interactions seem to be more attractive
than the Nijmegen soft-core potential NSC89 \cite{NSC89}.
The model FSS gives a large overbinding
since the $\hbox{}^1S_0$ $\Lambda N$ interaction is strongly
attractive. The phase-shift difference
of the $\hbox{}^1S_0$ and $\hbox{}^3S_1$ states
at $p_\Lambda \sim 200~\hbox{MeV}/c$ is $\delta(\hbox{}^1S_0)
-\delta(\hbox{}^3S_1) \sim 29^\circ$ for FSS,
while $\delta(\hbox{}^1S_0)-\delta(\hbox{}^3S_1)
\sim 7^\circ$ for fss2. Since the present fss2 result
is still slightly overbound, this difference should be somewhat
smaller in order to reproduce the correct experimental
value ${B_\Lambda}^{\rm exp}=130 \pm 50~\hbox{keV}$.
From the two results given by fss2 and FSS, we can extrapolate that,
in our quark model, the desired difference is $0^\circ \sim 2^\circ$.

In order to make sure that this extrapolation gives
a good estimation,
we modify the $\kappa$-meson mass of the model fss2
from the original value, $m_\kappa=936$ MeV \cite{fss2}, to 1,000 MeV,
and repeat the whole calculation. It is known that
this modification makes the $\hbox{}^1S_0$ $\Lambda N$ interaction
less attractive and the $\hbox{}^3S_1$ more attractive.
We obtain $B_\Lambda=145$ keV with $P_\Sigma=0.53\,\%$.
The effective range parameters of this modified fss2 interaction
are $a_s=-2.15$ fm, $r_s=3.05$ fm,
and $a_t=-1.80$ fm, $r_t=2.87$ fm. The phase-shift difference is
only $1.3^\circ$ and the total cross section
of the $\Lambda N$ scattering increases at most 10 mb at
$p_{\rm \Lambda}=100~\hbox{MeV}/c$ from 286 mb to 296 mb,
which is still within the experimental error bars.

It should be kept in mind that the effective range parameters
or the $S$-wave phase-shift values determined in this way is
very much model dependent, since the $B_\Lambda$ value is not
solely determined by these quantities. It depends on how higher
partial waves contribute and also on the details
of the $\Lambda N$-$\Sigma N$ coupling of a particular model.
A nice extrapolation shown here is based on the similarity
of the models fss2 and FSS, which have a common framework
for the quark sector and the effective meson-exchange potentials. 

\begin{table}[t]
\caption{$\hbox{}^1S_0$ and $\hbox{}^3S_1$ effective
range parameters of FSS \protect\cite{FSS}, fss2
\protect\cite{fss2,B8B8}, and NSC89 \protect\cite{NSC89}
$\Lambda N$ interactions ($\Lambda p$ for NSC89) and
the $\Lambda$ separation energies $B_\Lambda$ of the
hypertriton. The values for NSC89 are taken from Ref.~\protect\cite{No02}.
}
\label{table3}
\renewcommand{\tabcolsep}{0.5pc} 
\renewcommand{\arraystretch}{1.2} 
\begin{tabular}{@{}cccccc}
\hline
model & $a_s$ (fm) & $r_s$ (fm) & $a_t$ (fm) & $r_t$ (fm)
& $B_\Lambda$ (keV) \\
\hline
FSS & $-5.41$ & 2.26 & $-1.03$ & 4.20
& 878 \\
fss2 & $-2.59$ & 2.83 & $-1.60$ & 3.01
& 289 \\
NSC89 & $-2.59$ & 2.90 & $-1.38$ & 3.17
& 143 \\
\hline
\end{tabular}
\end{table}

Table \ref{table2} also shows that the expectation value
of the $NN$ Hamiltonian, $\varepsilon_{NN}$, determined
self-consistently is rather close
to the deuteron energy $-\varepsilon_d$, especially in fss2.
This feature is even marked if we decompose these energies
to the kinetic-energy and potential-energy contributions.
Table \ref{table4} shows this decomposition with respect
to fss2, FSS and NSC89. (For this comparison,
we use the definition of the kinetic-energy part of the deuteron
by $h_d=\langle \chi_d|h_{NN}|\chi_d \rangle/\langle \chi_d|
\chi_d \rangle$, where $\chi_d$ is the RGM relative wave function
between the neutron and the proton.)
In fss2, the kinetic-energy of the $NN$ subsystem
is 1.88 MeV larger than that of the deuteron,
which implies that the $NN$ subsystem
shrinks by the effect of the outer $\Lambda$-particle, 
in comparison with the deuteron in the free space.
In NSC89, this difference is even smaller; i.e., 1.18 MeV.
These results are consistent with the fact
that the hypertriton in NSC89 is more loosely
bound ($B_\Lambda=143$ keV \cite{No02}) than in fss2 (289 keV),
and the $\Lambda$-particle is very far apart from the $NN$ cluster.
The $\Sigma NN$ probability in NSC89
is $P_\Sigma=0.5\,\%$ \cite{MI95,No02}.
Table \ref{table4} also lists the kinetic-energy and
potential-energy decompositions for the averaged $YN$ expectation
value $\varepsilon_{YN}$ and the total energy $E$.
The kinetic energies of $\varepsilon_{YN}$ are much smaller
than those of $\varepsilon_{NN}$, which indicates that the
relative wave functions between the hyperon and the nucleon are
widely spread in the configuration space.
The comparison of the total-energy decomposition shows that
the wave functions of fss2 and NSC89 may be very similar.
A clear difference between fss2 and NSC89 appears
in the roles of higher partial waves. The energy increase
due to the higher partial waves than the $S$ and $D$ waves
is 91 keV in fss2 and 168 keV in FSS, respectively.
On the other hand, the results in Ref.\ \cite{MI95}
imply that this is only 20 - 30 keV in the case of NSC89.
This difference can originate from both
of the $NN$ and $YN$ interactions. Since the characteristics 
of the meson-theoretical $YN$ interactions in higher partial
waves are a priori unknown, more detailed analysis of the
fss2 results might shed light on the adequacy of the
quark-model baryon-baryon interactions.

\begin{table}[t]
\caption{Decomposition of the $NN$ and $YN$ expectation
values ($\varepsilon_{NN}$ and $\varepsilon_{YN}$), the
deuteron energy ($-\varepsilon_d$) and the total three-body
energy $E$ to the kinetic-energy and potential-energy
contributions. The unit is in MeV. 
The results for NSC89 are from \protect\cite{MI95}.}
\label{table4}
\renewcommand{\tabcolsep}{2.0pc} 
\renewcommand{\arraystretch}{1.2} 
\begin{tabular}{@{}cc}
\hline
model & $h_{NN}+V_{NN}=\varepsilon_{NN}$ \\
\hline
FSS   & $19.986-21.303=-1.317$ \\
fss2  & $19.376-21.032=-1.657$ \\
NSC89 & $20.48 -22.25 =-1.77$ \\
\hline
model & $h_d+V_d=-\varepsilon_d$ (deuteron) \\
\hline
FSS   & $16.982-19.238=-2.256$ \\
fss2  & $17.495-19.720=-2.225$ \\
NSC89 & $19.304-21.528=-2.224$ \\
\hline
model & $h_{YN}+V_{YN}=\varepsilon_{YN}$ \\
\hline
FSS   & $10.036-4.602=5.435$ \\
fss2  &  $8.071-2.671=5.401$ \\
NSC89 &  $7.44 -3.54 =3.90$ \\
\hline
model & $\langle H_0 \rangle+\langle V \rangle=E$ \\
\hline
FSS   & $27.372-30.506=-3.134$ \\
fss2  & $23.860-26.374=-2.514$ \\
NSC89 & $23.45 -25.79 =-2.34$ \\
\hline
\end{tabular}
\end{table}

\section{Summary}

In this study, we have carried out the Faddeev calculations,
using the recent quark-model $NN$ and $YN$ interactions,
FSS \cite{FSS} and fss2 \cite{fss2,B8B8}.
These are realistic interactions which describe
all the available $NN$ and $YN$ data, by incorporating the
effective meson-exchange potentials at the quark level.
Since these quark-model baryon-baryon interactions are formulated
in the RGM framework, they are non-local, energy-dependent,
and sometimes involve the Pauli-forbidden component
at the quark level.
The hypertriton is an appropriate place to investigate
the roles of the compact $SU_3$ $(11)_s$ component,
which is completely Pauli-forbidden
in the $\hbox{}^1S_0$ $\Lambda N$-$\Sigma N$ channel
coupling with the isospin $T=1/2$.
In order to deal with this off-shell effect of the quark-model
interaction precisely, we have formulated a new type
of the Faddeev equation which explicitly employs
the two-cluster RGM kernels \cite{TRGM,RED}.
The energy-dependence of the RGM kernel is self-consistently
treated, by calculating the expectation values of the
two-cluster Hamiltonian with respect to the obtained solutions
of the Faddeev equation \cite{BE9L}.
We have especially payed attention to how to extend the
microscopic description of the $\Lambda N$-$\Sigma N$ coupling
in the hypertriton system without spoiling the effect of the
Pauli-principle at the quark level. 
The present study is the second application of this formalism
to the few-baryon systems interacting via the quark-model
baryon-baryon interactions, following our previous one
to the triton system \cite{triton,PANIC02}.
The hypertriton is well suited to investigate the on-shell
properties of the $\Lambda N$ and $\Sigma N$ interactions,
since the hyperon is very far apart from the two-nucleon cluster.  

We have found that our quark-model interaction fss2 gives
a reasonable result for the hypertriton properties,
which is rather similar to the result of the Nijmegen
soft-core model NSC89 \cite{NSC89}.
The $\Lambda$ separation energy given by fss2
is $B_\Lambda=289$ keV,
which is a little too large in comparison with the experimental
value ${B_\Lambda}^{\rm exp}=130 \pm 50~\hbox{keV}$.
The admixture of the $\Sigma NN$ component
is $P_\Sigma=0.80\,\%$.
Modifying the $\kappa$-meson mass of fss2
from the original value, $m_\kappa=936$ MeV, to 1,000 MeV
leads to the almost correct $\Lambda$-separation energy
145 keV with $P_\Sigma=0.53\,\%$.
Unlike the NSC89 result, the effects of higher partial waves
up to the $G$ wave are rather important
in the quark-model $NN$ and $YN$ interactions.
If we use the dominant $S$-wave character
of the $\Lambda N$ interaction in the hypertriton system,
the $\hbox{}^1S_0$ $\Lambda N$ interaction of the model fss2 is
slightly too attractive. It is a future problem to investigate
whether or not a reduction in the $\hbox{}^1S_0$ attraction
like the modification $m_\kappa c^2=936$ MeV to 1,000 MeV
produces a favorable feature for the level spacing
of the $0^+$ and $1^+$ states of $\hbox{}^4_\Lambda
\hbox{H}$ and $\hbox{}^4_\Lambda \hbox{He}$ systems.
The fairly large charge symmetry breaking in these systems
is also an important issue to understand
the hyperon-nucleon interactions in detail.

\begin{acknowledgments}
This work was supported by Grants-in-Aid for Scientific
Research (C) from the Japan Society for the Promotion
of Science (JSPS) (Grant Nos.~15540270, 15540284 and 15540292).
\end{acknowledgments}

\appendix*

\section{Derivation of the three-cluster Faddeev equation
for the \mib{\Lambda NN}-\mib{\Sigma NN} system}

In this appendix, we discuss some essential points
to derive the Faddeev equation \eq{form6}
from the three-cluster equation \eq{form1}.
In the derivation, we extensively use the following
properties of the projection operator $P$ defined
in \eq{form5}:
\begin{eqnarray}
&({\rm i}) &~\Lambda_{3i} P=P \Lambda_{3i}=P \quad (i=1, 2),
\nonumber \\
& & ~\hbox{where}~\Lambda_{3i}=1-|u_{3i}\rangle \langle u_{3i}|
\ ,\nonumber \\
&({\rm ii}) &~\hbox{when}~\Psi \in [11],
~\forall~\langle u_{3i}|\Psi \rangle=0
\longleftrightarrow  P \Psi=\Psi\ ,\nonumber \\
&({\rm iii}) &~\hbox{when}~\Psi \in [11],~P \Psi=0 
\longleftrightarrow \exists~|f\rangle\ ,\nonumber \\
& & ~\Psi=|u_{31}f_2\rangle-|u_{32}f_1\rangle\ .
\label{a1}
\end{eqnarray}
Using the property (i), we can replace $V^{\rm RGM}_{3i}
(\varepsilon_{3i})$ in \eq{form1} by $v_{3i}(\varepsilon_{3i})
=\Lambda_{3i}V^{\rm RGM}_{3i}(\varepsilon_{3i})\Lambda_{3i}$
or 
\begin{eqnarray}
{\cal V}_{3i}(E,\varepsilon_{3i})
=(E-H_0)-\Lambda_{3i} (E-H_0) \Lambda_{3i}
+v_{3i}(\varepsilon_{3i})\ .\nonumber \\
\label{a2}
\end{eqnarray}
We further use the property (iii) for the whole equation and 
introduce the ansatz for the Faddeev components,
$P \Psi=\psi+(1-P_{12})\phi$, to derive a pair of equations
\begin{eqnarray}
\left[\,E-H_0-V^{\rm RGM}_{12}(\varepsilon_{12})\,\right] \psi
& = & V^{\rm RGM}_{12}(\varepsilon_{12})\left(1-P_{12}\right) \phi
\ ,\nonumber \\
\left[\,E-H_0-{\cal V}_{31}(E, \varepsilon_{31})\,\right]
\phi & = & {\cal V}_{31}(E, \varepsilon_{31})
\left(\psi-P_{12}\phi\right) \nonumber \\
& & +|u_{31}f_2\rangle\ .
\label{a3}
\end{eqnarray}
%
\noindent
In the second equation, we note that
\begin{eqnarray}
& & 
E-H_0-{\cal V}_{31}(E, \varepsilon_{31})
\nonumber \\
& & 
=\Lambda_{31}\left[E-H_0-V^{\rm RGM}_{3i}
(\varepsilon_{3i})\right] \Lambda_{31}\ ,
\label{a4}
\end{eqnarray}
and introduce the projected two-body Green function in the
three-body space, $G_{\Lambda_{31}}(E,\varepsilon_{31})$,
which satisfies
%
\begin{eqnarray}
& & G_{\Lambda_{31}}(E, \varepsilon_{31})
\,\Lambda_{31}\left[\,E-H_0-v_{31}(\varepsilon_{31})\,\right]
\Lambda_{31}=\Lambda_{31}\ .
\nonumber \\
\label{a5}
\end{eqnarray}
%
This can be easily constructed through
%
\begin{eqnarray}
& & G_{\Lambda_{31}}(E, \varepsilon_{31})
=G_{v_{31}}(E, \varepsilon_{31})
-G_{v_{31}}(E, \varepsilon_{31})|u_{31}\rangle
\nonumber \\
& & \times 
{1 \over \langle u_{31}|
G_{v_{31}}(E, \varepsilon_{31})|u_{31}\rangle}
\langle u_{31}|
G_{v_{31}}(E, \varepsilon_{31})\ ,
\label{a6}
\end{eqnarray}
by using the two-body Green function
$G_{v_{31}}(E, \varepsilon_{31})
=\left[E-H_0-v_{31}(\varepsilon_{31})+i0\right]^{-1}$ in
the three-body space.
The essential equation we need for deriving the full
Green function $G_{31}(E, \varepsilon_{31})
=\left[E-H_0-V^{\rm RGM}_{31}(\varepsilon_{31})+i0\right]^{-1}$
is the decomposition
%
\begin{eqnarray}
& &  E-H_0-V^{\rm RGM}_{31}(\varepsilon_{31})
=\left(E-\varepsilon_{31}-h_{\bar 2}-\Delta m_3\right)
\nonumber \\
& & \quad -\Lambda_{31}\left(E-\varepsilon_{31}
-h_{\bar 2}-\Delta m_3\right)\Lambda_{31}
\nonumber \\
& & \quad +\Lambda_{31}
\left[E-H_0-V^{\rm RGM}_{31}(\varepsilon_{31})\right]
\Lambda_{31} \nonumber \\
& & = |u_{31}\rangle \left(E-\varepsilon_{31}-h_{\bar 2}-\Delta m_3
\right) \langle u_{31}|
\nonumber \\
& & \quad +\Lambda_{31}
\left[E-H_0-V^{\rm RGM}_{31}(\varepsilon_{31})\right]
\Lambda_{31}\ , 
\label{a7}
\end{eqnarray}
but the last equality is actually not satisfied
since $\left(E-\varepsilon_{31}-h_{\bar 2}-\Delta m_3\right)$ is
channel dependent. This difficulty is avoided
by using $V^{\rm mod}_{\rm RGM}(\varepsilon)$ in \eq{form16},
in place of $V^{\rm RGM}_{31}(\varepsilon_{31})$.
In fact, we find that
%
\begin{eqnarray}
& & \left(E-\varepsilon_0-h_{\bar 2}-\Delta m_3\right)
-\left(h_{31}-h_0\right)
+\left(\Delta E_{\rm int}+\Delta \varepsilon \right.
\nonumber \\
& & \left.
+\Delta h_0\right)
=E-\varepsilon_{\Lambda N}-\left(h_{\bar 2}\right)_\Lambda
\label{a8}
\end{eqnarray}
is channel independent.
Here $\varepsilon_0$ and $\left(h_{31}-h_0\right)$ term appear
since $|u_{31}\rangle$ actually satisfies \eq{form10}
and not \eq{form3}.
This makes it possible to derive our basic relationship
\begin{widetext}
\begin{eqnarray}
G_0(E) T_{31}(E,\varepsilon)
& = & G_{31}(E, \varepsilon)V^{\rm mod}_{\rm RGM}(\varepsilon)
=G_{\Lambda_{31}}(E, \varepsilon)
{\cal V}_{31}(E, \varepsilon)
-|u_{31}\rangle \langle u_{31}|
+|u_{31}\rangle {1 \over 
E-\varepsilon_{\Lambda N}-\left(h_{\bar 2}\right)_\Lambda}
\langle u_{31}|(E-H_0) \nonumber \\
& = & G_0(E) \widetilde{T}_{31}(E, \varepsilon)
+|u_{31} \rangle {1 \over 
E-\varepsilon_{\Lambda N}-\left(h_{\bar 2}\right)_\Lambda}
\langle u_{31}|(E -H_0)\ ,
\label{a9}
\end{eqnarray}
\end{widetext}
where all the kernels are defined
by using $V^{\rm mod}_{\rm RGM}(\varepsilon)$.
From \eq{a9} we can easily prove the second Faddeev
equation \eq{form6} and the orthogonality
condition \eq{form11}.


\begin{thebibliography}{9}
\bibitem{FSS} Y. Fujiwara, C. Nakamoto, and Y. Suzuki,
Phys. Rev. Lett. {\bf 76}, 2242 (1996);
Phys. Rev. C {\bf 54}, 2180 (1996).
\bibitem{fss2} Y. Fujiwara, T. Fujita, M. Kohno, C. Nakamoto,
and Y. Suzuki, Phys. Rev. C {\bf 65}, 014002 (2002).
\bibitem{B8B8} Y. Fujiwara, M. Kohno, C. Nakamoto, and Y. Suzuki,
Phys. Rev. C {\bf 64}, 054001 (2001).
\bibitem{TRGM} Y. Fujiwara, H. Nemura, Y. Suzuki, K. Miyagawa,
and M. Kohno, Prog. Theor. Phys. {\bf 107}, 745 (2002).
\bibitem{RED} Y. Fujiwara, Y. Suzuki, K. Miyagawa, M. Kohno,
and H. Nemura, Prog. Theor. Phys. {\bf 107}, 993 (2002).
\bibitem{triton} Y. Fujiwara, K. Miyagawa, M. Kohno, Y. Suzuki,
and H. Nemura, Phys. Rev. C {\bf 66}, 021001(R) (2002).
\bibitem{PANIC02} Y. Fujiwara, K. Miyagawa, Y. Suzuki, M. Kohno,
and H. Nemura, Nucl. Phys. {\bf A721}, 983c (2003).
\bibitem{MA89} R. Machleidt, Adv. Nucl. Part. Phys.
{\bf 19}, 189 (1989).
\bibitem{No00} A. Nogga, H. Kamada, and W. Gl{\" o}ckle,
Phys. Rev. Lett. {\bf 85}, 944 (2000).
\bibitem{MI95} K. Miyagawa, H. Kamada, W. Gl{\"o}ckle, and V. Stoks,
Phys. Rev. C {\bf 51}, 2905 (1995).
\bibitem{MI00} K. Miyagawa, H. Kamada, W. Gl{\"o}ckle, H. Yamamura,
T. Mart, and C. Bennhold, Few-Body Systems Suppl.
{\bf 12}, 324 (2000).
\bibitem{No02} A. Nogga, H. Kamada, and W. Gl{\"o}ckle,
Phys. Rev. Lett. {\bf 88}, 172501 (2002);
A. Nogga, Ph. D. thesis, University of Bochum (2001).
\bibitem{NSC89} P. M. M. Maessen, Th. A. Rijken, and J. J. de Swart,
Phys. Rev. C {\bf 40}, 2226 (1989).
\bibitem{NSC97} Th. A. Rijken, V. G. J. Stoks, and Y. Yamamoto,
Phys. Rev. C {\bf 59}, 21 (1999).
\bibitem{GRGM} Y. Fujiwara, M. Kohno, C. Nakamoto, and Y. Suzuki,
Prog. Theor. Phys. {\bf 104}, 1025 (2000).
\bibitem{GL82a} W. Gl{\" o}ckle, G. Hasberg, and A.R. Neghabian,
Z. Phys. A {\bf 305}, 217 (1982).
\bibitem{BE9L} Y. Fujiwara, K. Miyagawa, M. Kohno, Y. Suzuki,
D. Baye, and J.-M. Sparenberg, KUNS-1910, nucl-th/0404071,
submitted to Phys. Rev. C.
\bibitem{AR96} See ARPACK homepage, http://www.caam.rice.edu/
software/ARPACK/
%
\end{thebibliography}
\end{document}